\documentclass[twocolumn,showpacs,preprintnumbers,amsmath,amssymb,floatfix]{revtex4}

\usepackage{graphicx}
\usepackage{epsfig} 
\usepackage{dcolumn}
\usepackage{bm}
\usepackage{epstopdf}
\usepackage{multirow}
\usepackage{amsmath}
\usepackage{booktabs}

\begin{document}

\preprint{APS/123-QED}

\title{Production of ultracold ground state LiRb molecules by photoassociation through a resonantly coupled state}

\author{I. C. Stevenson$^1$, D. Blasing$^2$, Y. P. Chen$^{2,1,3}$, and D. S. Elliott$^{1,2,3}$}
\affiliation{%
  $^1$School of Electrical and Computer Engineering and $^2$Department of Physics and Astronomy and
  $^3$Purdue Quantum Center\\ Purdue University, West Lafayette, IN  47907
}

\date{\today}

\begin{abstract}
	We report on a resonantly coupled 2(1) - 4 (1) photoassociation resonance in LiRb.  This pair of states displays interesting decay physics that hint at interference-like effects caused by two different decay paths.  We observe decay to predominantly $X \ ^1\Sigma^+ \ v=43$, with significant numbers of molecules produced in $X \ ^1\Sigma^+ \ v=2$ - 12.  This photoassociation resonance also produces $\sim$300 $X \ ^1\Sigma^+ \ v=0 \ J=0$ molecules/second.  Finally, we identify a stimulated Raman adiabatic passage transfer pathway from $v=43$ to $v=0$ that has the potential to produce up to $2 \times 10^5$ LiRb molecules/second in the $X \ ^1\Sigma^+ \ v=0 \ J=0$ state.
\end{abstract}

\maketitle

\section{Introduction}

Ultracold polar molecules are interesting for many reasons~\cite{demille2002quantum,ni2010dipolar,krems2009cold}.  In particular, their permanent electric dipole moment in the rovibronic ground state produces unique physical interactions~\cite{trefzger2009pair,capogrosso2010quantum,pupillo2008cold}.  To realize many of these goals, a continuous means of generating a high density sample of rovibronically-cold dipolar molecules would be very useful.  Currently, the most successful polar molecule experiments use an experimentally challenging pulsed technique.  Molecules are produced via magneto-association and transfered to the ro-vibronic ground state (that is vibrational, rotational and electronic ground state) with stimulated Raman adiabatic passage (STIRAP)~\cite{ni2008high}.  However, ground state generation by direct spontaneous emission after photoassociation (PA) has been observed previously in KRb~\cite{banerjee2012direct}, LiCs~\cite{deiglmayr2008formation}, RbCs~\cite{bruzewicz2014continuous} and NaCs~\cite{zabawa2011formation}, explored theoretically in Ref.~\cite{stwalley2009resonant}.  This method is experimentally simpler and continuous.  There have been several proposals for trapping these continuously generated ground state molecules~\cite{bruzewicz2014continuous,kleinert2007trapping} and several for increasing the phase-space density to be competitive with the pulsed production method~\cite{wakim2012luminorefrigeration,kobayashi2014prospects}.  For the continuous generation method to be successful, a PA resonance that produces large numbers of cold molecules, preferably in one vibrational state, is essential.

In this work, we investigate a newly-discovered photoassociation resonance, a mixed state consisting of resonantly coupled $2(1) - 4(1)$ long-range states in LiRb that produces large numbers of deeply bound molecules.  Similar resonantly-coupled states exist in other molecules~\cite{banerjee2012direct,zabawa2011formation,stwalley2009resonant} and spontaneous decay to the electronic $X \: ^1 \Sigma ^+$ ground state, including the lowest vibrational state, has been observed; we find this holds true for LiRb.  Molecules associated through this resonance decay to many vibrational levels of both the $X \: ^1 \Sigma ^+$ ground state and the $a \: ^3 \Sigma ^+$ lowest triplet state, with nearly half of these molecules decaying to $X \: ^1 \Sigma ^+ \ v=43$.  We use resonantly enhanced multiphoton ionization (REMPI) and a type of depletion spectroscopy to explore the production rate and distribution of the molecules; because of its strong decay to stable molecular states, we have found this PA resonance to be a very successful platform to study bound-to-bound transitions in LiRb~\cite{stevensonCE16}.  Finally, as part of this work we present the first observations of ultracold $X \ ^1\Sigma^+ \ v=0 \ J=0$ LiRb molecules formed at around 300 molecules/second and discuss plans for producing up to $2 \times 10^5$ ro-vibronic ground states molecules/second.  The relevant potential energy curves (PEC) for these measurements are in Fig.~\ref{fig:2(1)4(1)PA}. 

\begin{table}
	\centering
	\begin{tabular}{cc}
		\hline \hline
		Hund's case (a) or (b) & Hund's case (c)\\
		\midrule[1.pt]
		$A \: ^1 \Sigma ^+$ & $2(0^+)$\\
		\multirow{2}{*}{$c \: ^3 \Sigma ^+$} & $2(0^-)$\\
		& $2(1)$\\
		$B \: ^1 \Pi$ & $4(1)$\\
		\hline \hline
	\end{tabular}
	\caption{Correspondence between Hund's cases (a) and (c) in LiRb.  We use Hund's case (c) labeling for our PA states and Hund's case (a) for bound to bound transitions.  Note that $B \: ^1 \Pi \ v^\prime=20$ and $4(1) \ v=-16$ denote the same state.}
	\label{tab:HundsCase}
\end{table}

\section{Experimental parameters} \label{sec:ExpParams}

\begin{figure}[t]
	\includegraphics[width=8.6cm]{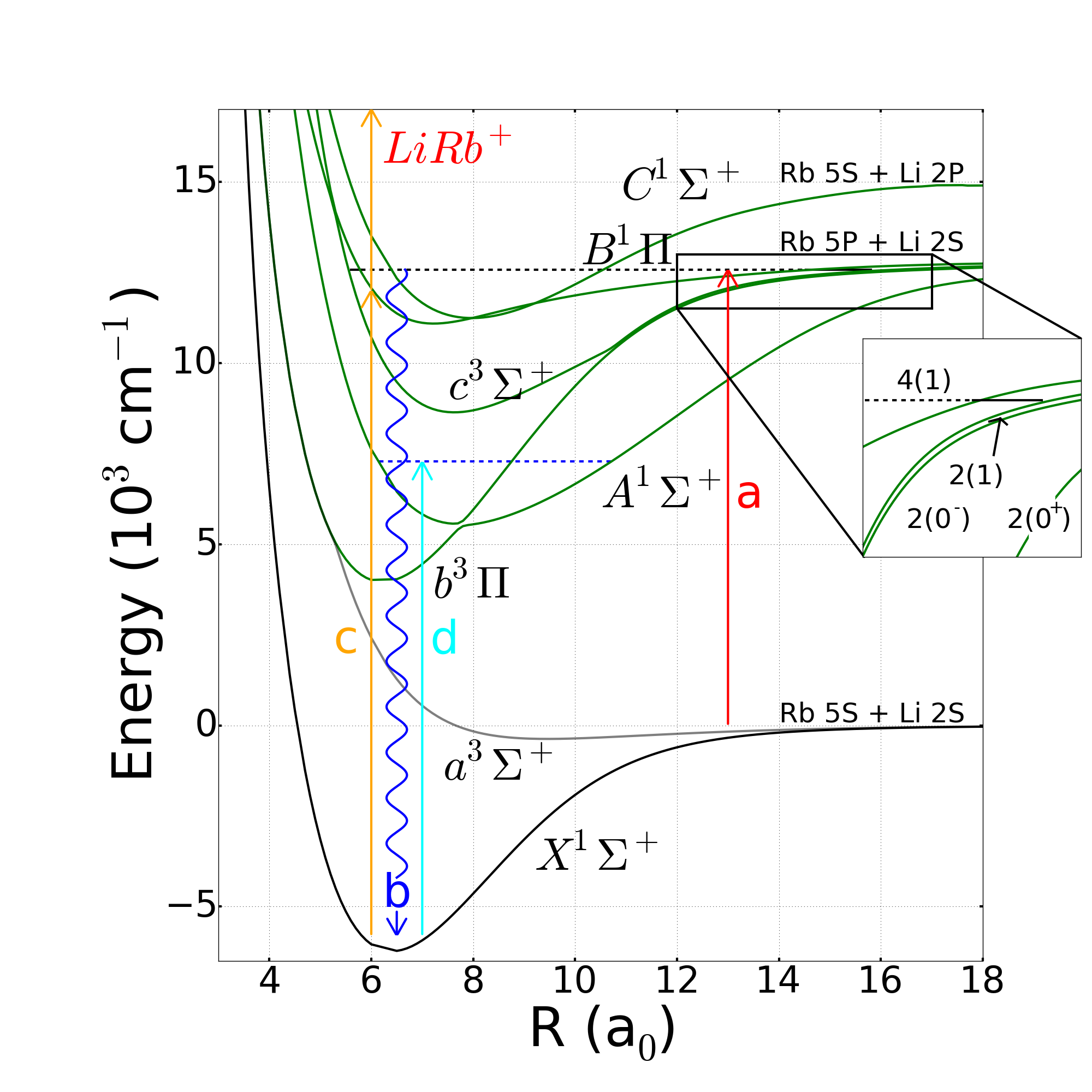}\\
	\caption{(Color online) Low-lying PEC diagram for the LiRb molecule from Ref.~\cite{korek2009theoretical}.  Vertical lines show various optical transitions, including {\bf (a)} photoassociation of molecules below the Rb D$_1$ asymptote, at frequency $\nu_{a}$; {\bf (b)} spontaneous decay of excited state molecules leading to $X \: ^1 \Sigma ^+$ (and  $a \: ^3 \Sigma ^+$); {\bf (c)} REMPI or RE2PI to ionize LiRb molecules, at frequency $\nu_{c}$; and {\bf (d)} state-selective excitation to deplete the REMPI signal, at frequency $\nu_{d}$.  The black and blue dashed lines respectively represents our 2(1) - 4(1) mixed states and our depletion state.  The inset shows an expanded view of the long-range potentials near the asymptote.  Table~\ref{tab:HundsCase} shows the correlation between long-range and short-range labeling.}
	\label{fig:2(1)4(1)PA}
\end{figure}

The details of our experimental apparatus are described elsewhere~\cite{dutta2014interspecies} so we provide only a brief summary.  We work out of a dual-species MOT, $\lesssim$1 mK in temperature and 1 mm in diameter~\cite{altaf2015formation}, trapping $\sim 5 \times 10^7$ $^7$Li atoms and $\sim 10^8$ $^{85}$Rb atoms.  Our Rb MOT is a spatial dark spot MOT~\cite{ketterle1993high}.  We photoassociate Li and Rb atoms using the output of either a 300 mW cw Ti:Sapphire laser, or a 150 mW external-cavity diode laser (ECDL) tuned below the Rb $D_1$ asymptote at 795 nm.  The beams are collimated to a $1/{e^{2}}$ diameter of 0.7 mm  at the center of the MOTs. After photoassociation, the LiRb molecules rapidly decay to the $X \ ^1\Sigma^+$ or $a \ ^3\Sigma^+$ electronic states.  We then ionize these molecules with the REMPI process driven by a Nd:YAG pumped pulsed dye laser, tunable in the wavelength range between 550 - 583 nm (18150 - 17150 cm$^{-1}$).  The repetition rate of this laser is 10 Hz, and it delivers $\sim$3 mJ/pulse to the MOT region in a 4 mm diameter beam.  When one-photon of this laser is resonant with a transition from an initial state (populated by spontaneous decay from the PA state) to an intermediate bound state (usually one of the vibrational levels of the $B \: ^1 \Pi$, $D \: ^1 \Pi$, $f \: ^3 \Pi$, or $g \ ^3\Sigma^+$ electronic potentials, not shown in Fig.~\ref{fig:2(1)4(1)PA}), then the molecule may absorb two photons and ionize.  We detect and count LiRb$^+$ ions using a time-of-flight spectrometer and a microchannel plate detector.  To observe ions from deeply bound states we use a two-color variant REMPI called RE2PI.  In RE2PI we reduced the power of dye laser pulse, and used the energetically higher photon from the 532 nm second-harmonic output of the Nd:YAG as the second photon.  (Notation: $v^{\prime \prime}$ and $J^{\prime \prime}$ denote the vibrational and rotational levels of the $X \: ^1 \Sigma ^+$ and $a \ ^3\Sigma^+$ states, $v$ and $J$ denote the vibrational and rotational levels of the electronically excited states driven by PA resonances (and for these vibrational numbers, we count down from the asymptote using negative integers), and $v^{\prime}$ and $J^{\prime}$ denote vibrational and rotational labeling of other excited electronic states.) 

\section{Resonant Coupling Evidence}\label{sec:ResCoup}
 
\begin{figure*}[t] 
	\includegraphics[width=8.6cm]{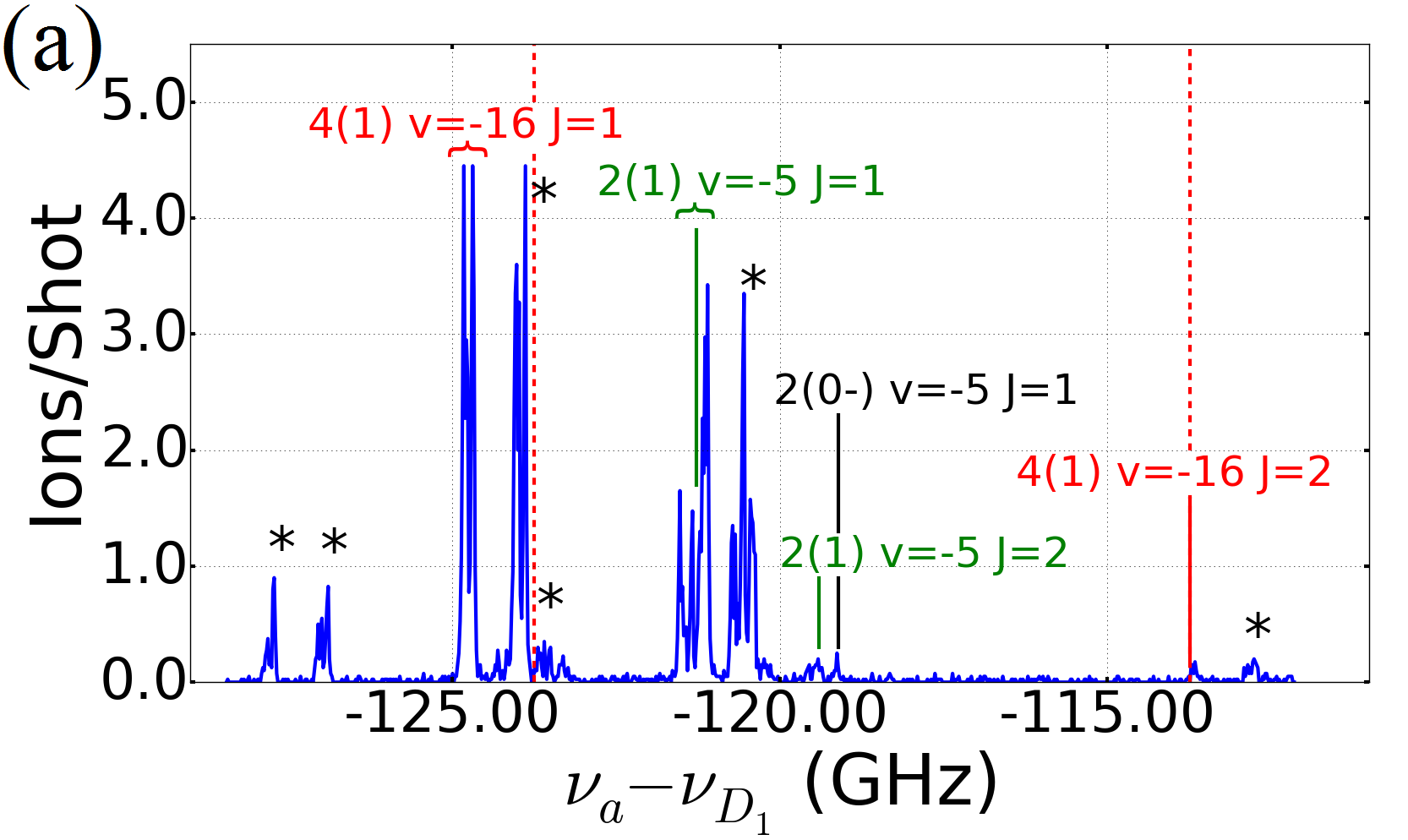}
	\includegraphics[width=8.6cm]{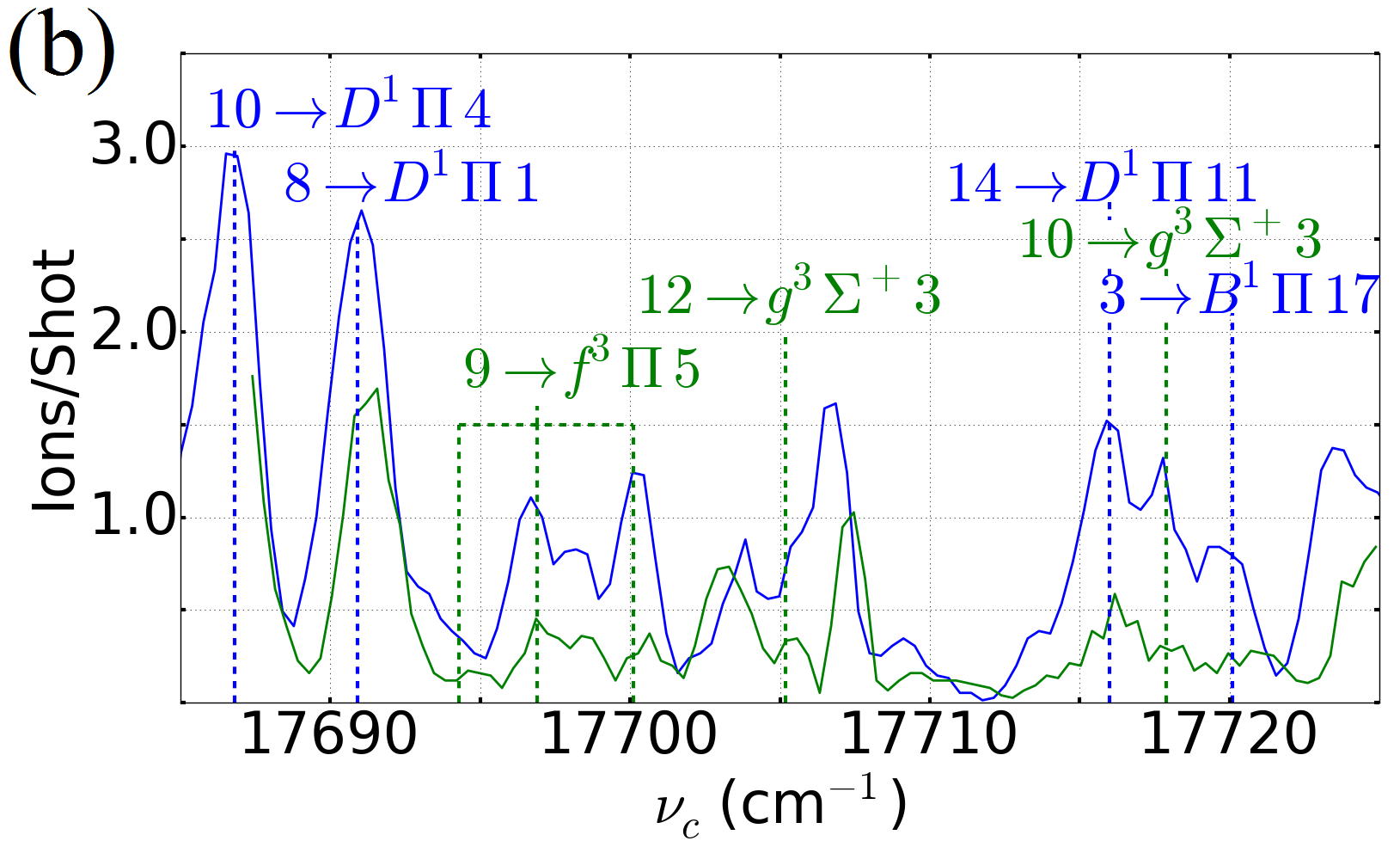}\\
	\caption{(Color online) Evidence of $2(1) - 4(1)$ mixing.  {\bf (a)} PA spectrum of the $2(1) - 4(1)$ mixed states.  The REMPI laser frequency was held fixed at 17,686.8 cm$^{-1}$ on the $D \ ^1\Pi \ v^\prime=4 \leftarrow v^{\prime \prime}=10$ transition.  $\ast$  Denotes hyperfine echoes.  New 4(1) resonances as well as previously observed 2(1) and 2(0$^-$) are labeled~\cite{altaf2015formation}.  Dashed lines show the expected position of the 4(1) resonance based on previous work~\cite{dutta2011laser}.  {\bf (b)} REMPI spectra from PA through the 4(1) and 2(1) states; (blue) 4(1) $v=-16 \ J=1$, (green) 2(1) $v=-5 \ J=1$.  The blue dashed lines represent transitions out of $X \ ^1\Sigma^+$ and the green dashed lines are transitions out of $a \ ^3\Sigma^+$; labeling is $v^{\prime \prime}$ $\rightarrow$ (excited state) $v^\prime$.}
	\label{fig:4(1)and2(1)REMPI}
\end{figure*}

In Fig.~\ref{fig:4(1)and2(1)REMPI}(a) we show the PA spectrum of the $2(1) - 4(1)$ mixed states near 122 GHz below the D$_1$ asymptote.  For this spectrum the REMPI laser frequency, $\nu_c$, was fixed on the $D \ ^1\Pi \ v^{\prime}=4 \leftarrow X \ ^1\Sigma^+ \ v^{\prime \prime} = 10$ transition and the PA laser frequency, $\nu_a$, was scanned.  The $4(1) \ v=-16 \ J=1$ and $2(1) \ v=-5 \ J=1$ states in this spectrum are coupled~\cite{facts1}, forming mixed states that possess characteristics of each~\cite{facts1.5}, namely, the good PA amplitude of a $2(1)$ state and the deep decay path of a $4(1)$ state. Coupling and mixing between vibrational levels of different electronic states can occur when states with the same rotational number $J$ and angular momentum $\Omega$ lie close energetically~\cite{banerjee2012direct,herzberg1951molecular}.  $\Omega$ is the projection of the total angular momentum (excluding nuclear spin) onto the internuclear axis.  Fig.~\ref{fig:4(1)and2(1)REMPI}(a) contains several hyperfine echoes, labeled with an $\ast$.  These are weaker PA lines that originate from population in our MOTs not in our main hyperfine component (i.e. $^7$Li population in F=1 and $^{85}$Rb population in F=3).

There are three features in Fig.~\ref{fig:4(1)and2(1)REMPI} that provide evidence of resonant coupling of the $2(1)$ and $4(1)$ states. 

\textbf{(a)} A $2(1)$ component is indicated by the large PA amplitude of the $4(1) \ v=-16$ state.  We previously explored the $v=-3$, $-4$, and $-5$ lines of the $4(1) $ state~\cite{lorenz2014formation}, and found that the PA amplitude for any vibrational levels more deeply bound than $v=-6$ had vanished.  The enhanced PA amplitude for $4(1) \ v=-16$, with no visible PA at $v=-15$ and $v=-17$, is an indication of mixing with a $2(1)$ state.

\textbf{(b)} The frequency spacing between the $J=1$ and $J=2$ lines of the $4(1) \ v=-16$ state in Fig.~\ref{fig:4(1)and2(1)REMPI}(a) is increased by $\sim$1 GHz from the spacing expected based on earlier spectroscopy of LiRb~\cite{dutta2011laser}.  We have marked the approximate expected locations of the $J=1$ and $2$ lines with vertical dashed red lines using $B_v = 2.50$ GHz. 

\textbf{(c)} Further evidence of $2(1) - 4(1)$ mixing is given in the REMPI spectra shown in Fig.~\ref{fig:4(1)and2(1)REMPI}(b).  These provide an indirect measure of the relative spontaneous decay paths after photoassociation. We recorded these spectra by tuning the PA laser to the $4(1) \ v=-16 \ J=1$ line (blue spectrum) or the $2(1) \ v = -5 \ J=1$ (green spectrum).  We have identified and labeled the lines in these spectra, and found population in vibrational levels of the $X \ ^1\Sigma^+$ and $a \ ^3\Sigma^+$ electronic potentials.  Previously we observed that other $2(1)$ states decay only to $a \ ^3\Sigma^+$ states, while other $4(1)$ states decay solely to $X \ ^1\Sigma^+$ states~\cite{altaf2015formation,lorenz2014formation}.  However, the $4(1) \ v=-16 \ J=1$ and the $2(1) \ v=-5 \ J=1$ states decay to both $X \ ^1\Sigma^+$ and $a \ ^3\Sigma^+$, providing further evidence of their coupling.

Our current and previous spectroscopy on $4(1) \ v=-16$ state allows us to analyze two interesting spectroscopic quantities of LiRb.  First, this leads us to a new, higher precision determination of the well depth of the  $X \ ^1\Sigma^+$ potential.  We use the transition frequency of the $B \ ^1\Pi \ v^{\prime} = 20 \leftarrow X \ ^1\Sigma^+ \ v^{\prime \prime} = 0$ transition from Ref.~\cite{dutta2011laser}, the PA frequency of the $4(1) \ v=-16 \ J = 2$ peak in Fig.~\ref{fig:4(1)and2(1)REMPI}(a) at -113.6 GHz, the Rb atomic D$_1$ transition frequency of 12579.00 cm$^{-1}$~\cite{steck85rubidium} ($F^{\prime}=2 \leftarrow F^{\prime \prime}=2$), and the ground state vibrational spacing from~\cite{dutta2011laser}.  This allows us to report a $X \ ^1\Sigma^+$ well depth of 5928.08 (0.02) cm$^{-1}$ relative to the $^{85}$Rb 5s $^2$S$_{1/2}$ F=2 + $^{7}$Li 2s $^2$S$_{1/2}$ F=1 asymptote.  This determination agrees with, but is of much greater precision than, the previous value of 5927.9 (4.0) cm$^{-1}$~\cite{ivanova2011x}.

\begin{figure}[t]
	\includegraphics[width=8.6cm]{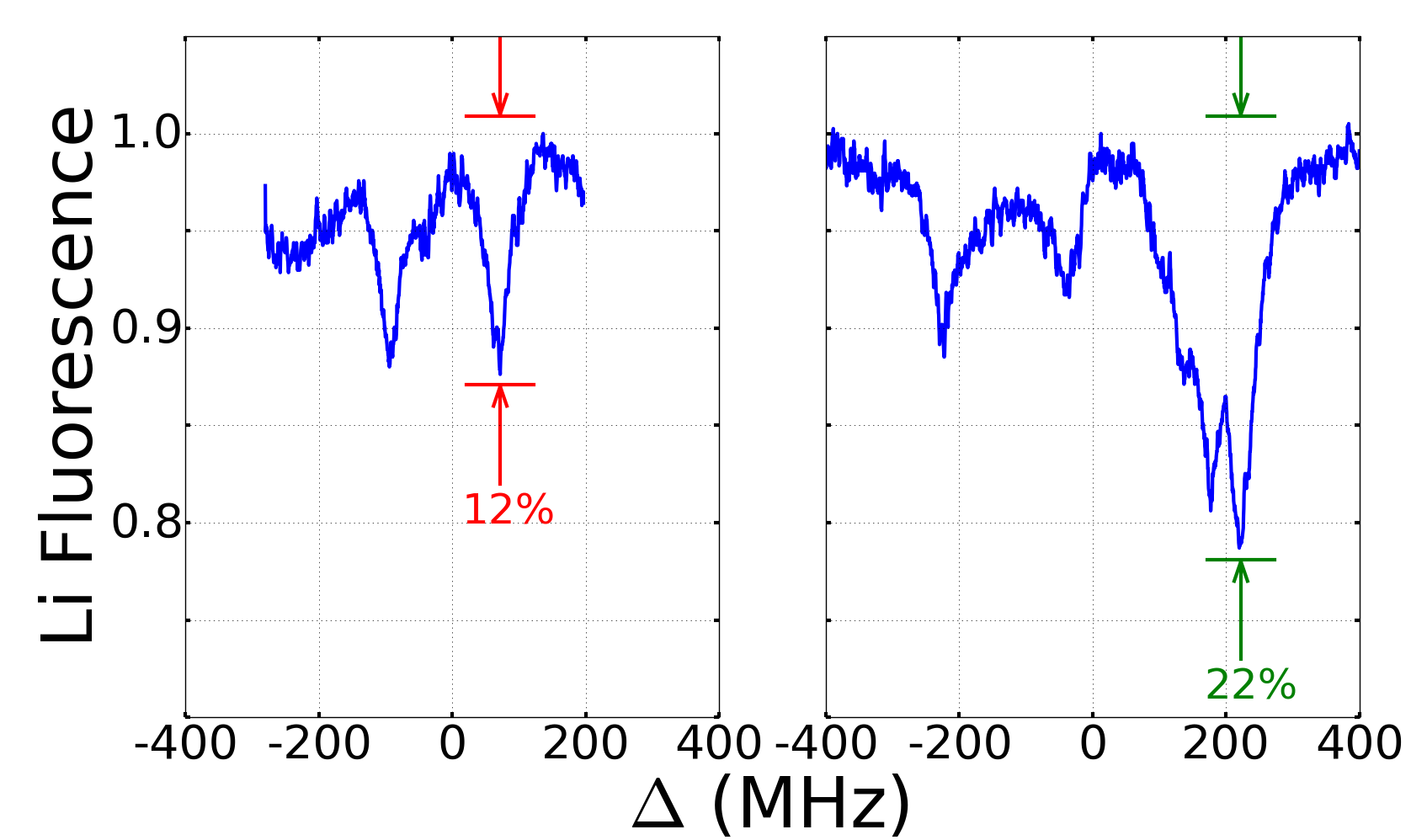}\\
	\caption{(Color online) Trap loss spectroscopy of $4(1) \ v=-16 \ J=1$ (left) and $2(1) \ v=-5 \ J=1$ (right) PA resonances.  Detuning is relative to the assigned PA line center.  Notice the relative strength favoring the 2(1) resonance in contrast to the REMPI data in Fig.~\ref{fig:4(1)and2(1)REMPI}(a).  More details on our trap loss experiments can be found in Ref.~\cite{dutta2014formation}.}
	\label{fig:trapLoss}
\end{figure}

We can also use this PA spectrum to estimate the admixture coefficients of the $2(1) - 4(1) \ J=1$ states.  We write the mixed states $| \Psi_- \rangle$ (primarily $4(1) \ J=1$) and $| \Psi_+ \rangle$ (primarily $2(1) \ J=1$) as
\begin{equation}
| \Psi_- \rangle = c | \Psi_{4(1)} \rangle - d | \Psi_{2(1)} \rangle \text{, with energy } E_-,
\label{eq:mixedStates-}
\end{equation}
and 
\begin{equation} 
| \Psi_+ \rangle = d | \Psi_{4(1)} \rangle + c | \Psi_{2(1)} \rangle \text{, with energy } E_+,
\label{eq:mixedStates+}
\end{equation}
where $| \Psi_{4(1)} \rangle$ and $| \Psi_{2(1)} \rangle$ are the bare (unmixed) states with energies $E_{4(1)}$ and $E_{2(1)}$, respectively.  We refer the reader to the treatment of resonantly coupled rotational states in Ref.~\cite{herzberg1951molecular}.  We measure the frequency difference of the perturbed states to be $(E_+ - E_-)/h = 3.5$ (0.1) GHz, as shown in the PA spectra of Fig.~\ref{fig:4(1)and2(1)REMPI}(a).  As discussed earlier, we know the unperturbed energy of the $4(1)$ state, marked with a red dashed line in Fig.~\ref{fig:4(1)and2(1)REMPI}(a).  Assuming an equal but opposite shift in the $2(1)$ energy, we find the energy difference of the unperturbed states to be $(E_{2(1)} - E_{4(1)})/h = 1.5$ (0.2) GHz.  Following the treatment of Ref.~\cite{herzberg1951molecular}, we can use the perturbed and unperturbed energy spacings to derive the coupling between the states, $V_{\rm int}$, by diagonalizing a simple 2x2 matrix,
\[ \begin{vmatrix}
E_{2(1)}-E & V_{\rm int} \\
V_{\rm int} & E_{4(1)}-E
\end{vmatrix}=0. \]
This will have solutions 
\begin{equation}
E_{\pm} = \frac{1}{2}(E_{2(1)}+E_{4(1)}) \pm \frac{1}{2}\sqrt{4 |V_{\rm int}|^2+\delta^2}.
\end{equation}  Re-solving this for the state coupling yields $V_{\rm int} = \frac{1}{2}\sqrt{(E_+-E_-)^2-\delta^2} = 1.6$ (0.2) GHz.  Additionally we can find the eigenstates of our diagonalized matrix to find the admixture coefficients $|c| = 0.84$ (0.09) and $|d| = 0.53$ (0.05), consistent with our earlier assertion of strong mixing between the states.  
\section{Relative Decay Paths}\label{sec:relativeDecay}
\begin{table*}
	\begin{tabular}{cccccccccccc}
		\hline \hline
		&   $\frac{R^-_{CM}}{R^+_{CM}}$   &   $\frac{B^+(v^{\prime \prime})}{B^-(v^{\prime \prime}) }$ & $\mathcal{R}(v^{\prime \prime})$ & $c^2 \ \mathcal{R}^2(v^{\prime \prime})$ & $2cd \mathcal{R}(v^{\prime \prime})$ & $d^2$ & $B^+(v^{\prime \prime})$ & $d^2 \ \mathcal{R}^2(v^{\prime \prime})$ & $-2cd \mathcal{R}(v^{\prime \prime})$ & $c^2$ & $B^-(v^{\prime \prime})$   \\ \midrule[1.pt]
		$X \ ^1\Sigma^+ \ v^{\prime \prime}=2$ & 2.5 (0.1) & 4.6  & -0.13 (0.05) & 0.01 & -0.12  & 0.29 & 0.18  & 0.00  & 0.12  & 0.71 & 0.83 \\
		$a \ ^3\Sigma^+ \ v^{\prime \prime}=7$ & 1.5 (0.1) &  2.8 & -0.02 (0.05) & 0.00 &  -0.02 & 0.29 & 0.27 & 0.00 & 0.02 & 0.71 & 0.73 \\
		$X \ ^1\Sigma^+ \ v^{\prime \prime}=43$ & 2.2 (0.3) & 4.0 & -0.10 (0.05) & 0.01 & -0.09 & 0.29 & 0.20 & 0.00 & 0.09 & 0.71 & 0.80 \\ \hline \hline
	\end{tabular}
	\caption{Enhancement of decay of $| \Psi_{-} \rangle$ across several vibrational levels.}\label{table:BR}
\end{table*}
To supplement the PA spectra discussed above, we have also observed the $4(1) - 2(1) \ J=1$ mixed resonance through trap loss measurements.  We show these spectra in Fig.~\ref{fig:trapLoss}.  In contrast to the REMPI signal, which indicates formation of a stable molecule in a specific vibrational level of the $X ^1 \Sigma^+$ or $a ^3 \Sigma^+$ state, a dip in the trap fluorescence signal results when the PA laser associates a molecule, regardless of whether that molecule decays to a stable state or to a pair of free atoms.  An interesting feature of these spectra is that the $|\psi_+ \rangle$ PA resonance, whose primary constituent is the bare $2(1) \ v=-5 \ J=1$ state, is stronger in trap loss (22\% trap loss) than the $|\psi_- \rangle$ line (12\% trap loss), while the $| \psi_- \rangle$ resonance is stronger than $|\psi_+ \rangle$ in the REMPI spectra.  In the following, we apply a simple model to characterize this.  

Using the fractional losses of 22\% and 12\%, we calculate that the PA rate to the $2(1) \ v=-5 \ J=1$ state, $| \Psi_+ \rangle$, is $R^+_{PA} = 1.9 \ (0.8) \times 10^{6}$ molecules/second and the PA rate to the $4(1) \ v=-16 \ J=1$ state, $| \Psi_- \rangle$, is $ R^-_{PA} = 1.1 \ (0.5) \times 10^{6} $
molecules/second.  The ratio of PA strengths, $R^+_{PA} / R^-_{PA}$, is 1.83 (0.15). (Most of the error in the individual PA rates is correlated, so the uncertainty in this relative measurement is much less than the quadrature sum of the individual uncertainties.)  

We can calculate the PA rate using a quantum perturbative framework~\cite{pillet1997photoassociation}
\begin{equation}
R_{PA} = \left( \frac{3\lambda_{th}^2}{2\pi} \right)^{\frac{3}{2}} \frac{h}{2} n_{Li} n_{Rb} A_\Omega |\langle \Psi_S | \vec{d} | \Psi_{PA} \rangle |^2.
\label{eq:definitionsPA}
\end{equation}
In Eq.~(\ref{eq:definitionsPA}), $| \Psi_S \rangle$ is the scattering wavefunction, $\lambda_{th}$ is the thermal de Broglie wavelength, $n_{Li}$ and $n_{Rb}$ are the atom densities and $A_\Omega$ are the radial factors for the two states and laser polarization.  $| \Psi_{PA} \rangle$ is either $| \Psi_- \rangle$ or $| \Psi_+ \rangle$, as given by Eqs.~({\ref{eq:mixedStates-}) or (\ref{eq:mixedStates+}).  Since none of the other nearby $4(1)$ vibrational levels have been observed, we assert that the cross section for PA to the deeply bound unperturbed $| \Psi_{4(1)} \rangle$ state is very small.  The other various factors in Eq.~(\ref{eq:definitionsPA}) are the same for $| \Psi_+ \rangle$ and $| \Psi_- \rangle$, so the relative PA rates are related approximately through
\begin{equation}
\frac{R^+_{PA}}{R^-_{PA}}=
\frac{| \langle \Psi_{S} | \vec{d} | \Psi_{+} \rangle |^2}{| \langle \Psi_{S} | \vec{d} | \Psi_{-} \rangle |^2} \simeq \frac{c^2 | \langle \Psi_{S} | \vec{d} | \Psi_{2(1)} \rangle |^2}{d^2 | \langle \Psi_{S} | \vec{d} | \Psi_{2(1)} \rangle |^2} \simeq \frac{c^2}{d^2}
\label{eq:paRatio}
\end{equation}
Using $|c| = 0.84$ and $|d|=0.53$ that we determined in Sec.~\ref{sec:ResCoup}, we estimate that the ratio of the trap loss peaks should be $\frac{c^2}{d^2} =$ 2.5 (0.6), which is indeed consistent with our trap loss data.  
	
We next examine the branching ratios for decay of excited state molecules to stable molecules, whose wavefunctions we designate  $| \Psi_{v^{\prime \prime}} \rangle$.  We write these as
\begin{multline}\label{eq:BRplus}
B^+(v^{\prime \prime}) =  |\langle \Psi_{v^{\prime \prime}} | \vec{d} | \Psi_{+} \rangle |^2 = c^2 |\langle \Psi_{v^{\prime \prime}} | \vec{d} | \Psi_{2(1)} \rangle |^2 \\ +  2 c d | \langle \Psi_{2(1)} | \vec{d} | \Psi_{v^{\prime \prime}} \rangle \langle \Psi_{v^{\prime \prime}} | \vec{d} | \Psi_{4(1)} \rangle  | \\ + d^2 | \langle \Psi_{v^{\prime \prime}} | \vec{d} | \Psi_{4(1)} \rangle |^2 ,
\end{multline}
and 
\begin{multline}\label{eq:BRminus}
B^-(v^{\prime \prime}) =  |\langle \Psi_{v^{\prime \prime}} | \vec{d} | \Psi_{-} \rangle |^2 = d^2 |\langle \Psi_{v^{\prime \prime}} | \vec{d} | \Psi_{2(1)} \rangle |^2 \\ -  2 c d | \langle \Psi_{2(1)} | \vec{d} | \Psi_{v^{\prime \prime}} \rangle \langle \Psi_{v^{\prime \prime}} | \vec{d} | \Psi_{4(1)} \rangle  | \\ + c^2 | \langle \Psi_{v^{\prime \prime}} | \vec{d} | \Psi_{4(1)} \rangle |^2 ,
\end{multline}
where $ \langle \Psi_{v^{\prime \prime}} | \vec{d} | \Psi_{2(1)} \rangle $ and $ \langle \Psi_{v^{\prime \prime}} | \vec{d} | \Psi_{4(1)} \rangle $ are the dipole transition matrix elements connecting the $v^{\prime \prime}$ vibrational state of $X \ ^1\Sigma^+$ or $a \ ^3\Sigma^+$ to the bare $2(1) \ v=-5 $ and $4(1) \ v=-16$ states.  Not shown in these equations is an overall normalization factor.  The ratio of these branching ratios becomes
\begin{equation}
\frac{B^+(v^{\prime \prime})}{B^-(v^{\prime \prime})}  \simeq \frac{c^2 \ \mathcal{R}^2(v^{\prime \prime}) + 2 c d \mathcal{R}(v^{\prime \prime}) + d^2 }{d^2 \ \mathcal{R}^2(v^{\prime \prime}) - 2 c d \mathcal{R}(v^{\prime \prime}) + c^2},
\label{eq:branchRatioExpanded}
\end{equation}
where $\mathcal{R}(v^{\prime \prime}) = \langle \Psi_{v^{\prime \prime}} | \vec{d} | \Psi_{2(1)} \rangle  / \langle \Psi_{v^{\prime \prime}} | \vec{d} | \Psi_{4(1)} \rangle $ is the ratio of the dipole transition matrix elements.

We determine this ratio $B^+(v^{\prime \prime})/B^-(v^{\prime \prime})$ for individual final states from the REMPI and trap loss spectra, and give three examples in Table~\ref{table:BR}.  $R^-_{CM}$ and $R^+_{CM}$ (column 2) are the cold molecule formation rates observed in REMPI in Fig.~\ref{fig:4(1)and2(1)REMPI}(a) when the PA laser is tuned to the $ | \Psi^- \rangle $ or $| \Psi^+ \rangle $ peak.  There we observed that approximately twice as many molecules were formed in deeply bound singlet states when PAing to $| \Psi_{-} \rangle$ compared to $| \Psi_{+} \rangle$ (this is difficult to see in Fig.~\ref{fig:4(1)and2(1)REMPI}(a) because of saturation effects).  The cold molecule formation rate $R_{CM}$ is related to the PA rate $R_{PA}$ through $R_{CM}=R_{PA} \times B(v^{\prime \prime})$. Using $R^+_{PA} / R^-_{PA}$ = 1.83, as discussed above, we determined the relative branching ratios $B^{+}(v^{\prime \prime})/B^{-}(v^{\prime \prime})$ for spontaneous decay to one of the vibrational states of $X \ ^1\Sigma^+$ or $a \ ^3\Sigma^+$ (column 3 in this table).  We solve Eq.~(\ref{eq:branchRatioExpanded}) for the ratio of transition moments $\mathcal{R}(v^{\prime \prime})$, using $|c/d|^2 = 2.5$, and list these values in Table~\ref{table:BR} as well.  We show here only one of the solutions for $\mathcal{R}(v^{\prime \prime})$.  The other root is a value of order 2, which seems unreasonable since other unmixed $2 (1)$ resonances lead to relatively small numbers of stable molecules.  
	
Finally, we examine the branching ratios $B^+(v^{\prime \prime})$ and $B^-(v^{\prime \prime})$, as computed using Eqs.~(\ref{eq:BRplus}) and (\ref{eq:BRminus}). We show each of the terms within these equations individually in Table~\ref{table:BR}.  Recall that we have omitted an overall normalization in these terms, as we are only examining their relative sizes.  There are some notable features to these results.  First, in any of the examples shown, for either of the PA resonances, the $|\langle \Psi_{v^{\prime \prime}} | \vec{d} | \Psi_{4(1)} \rangle |^2$ contribution is the most significant, while the $|\langle \Psi_{v^{\prime \prime}} | \vec{d} | \Psi_{2(1)} \rangle |^2$ contribution is insignificantly small.  Second, the cross term $2cd \mathcal{R}(v^{\prime \prime})$ makes a strong (secondary) contribution to each term.  This cross term contribution is of opposite sign for the two PA resonances.  That is, if it adds to the branching ratio of the $\Psi_-$ resonance, it diminishes the branching ratio of the $\Psi_+$ resonance.  Finally, observing the relative magnitudes of these terms, we note that the cross term is approximately half the magnitude of the strong $|\langle \Psi_{v^{\prime \prime}} | \vec{d} | \Psi_{4(1)} \rangle |^2$ contribution for the $\Psi_+$ resonance, strongly reducing this branching ratio.  This is consistent with our observations, in which we note very low generation of $X \ ^1\Sigma^+$ or $a \ ^3\Sigma^+$ molecules when tuned to the $| \Psi_+ \rangle$ PA peak.

\begin{figure}[b]
	\includegraphics[width=8.6cm]{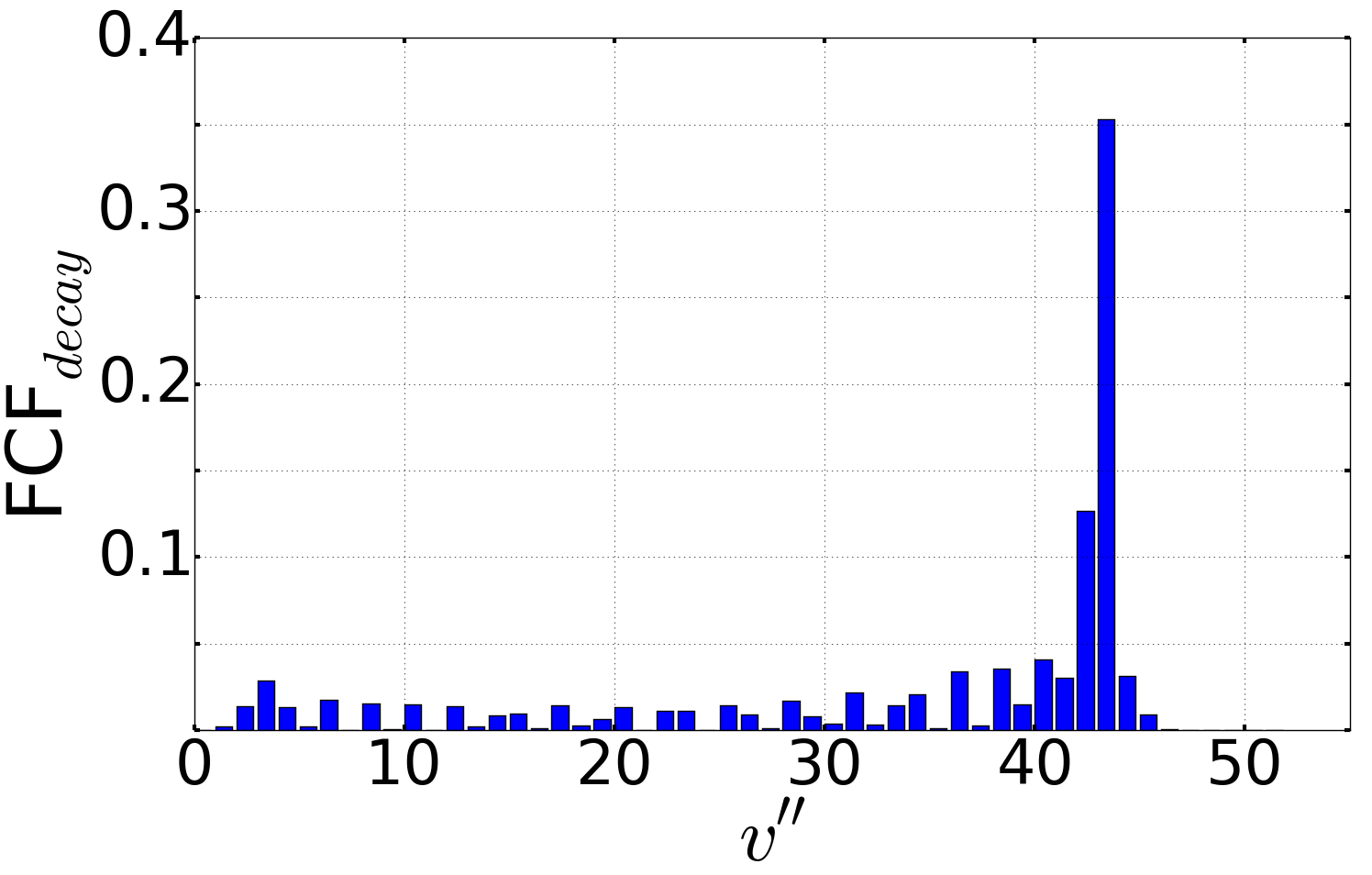}
	\caption{Calculated Franck - Condon factors using PECs from Ref.~\cite{ivanova2011x} with assistance from LEVEL 8.0~\cite{le2016level} for the decay path from the PA state $B \ ^1\Pi \ v^{\prime}=20$ ($4(1) \ v=-16$) to $X \ ^1\Sigma^+ \ v^{\prime \prime}$.  Notice the maximum, $v^{\prime \prime}=43$; no transitions from $v^{\prime \prime}=43$ show up in our REMPI data taken with the R590 dye.  All transitions observed can be assigned to $X \ ^1\Sigma^+ \ v^{\prime \prime}=0$ - 20 and $a \ ^3\Sigma^+ \ v^{\prime \prime}=6$ - 13.}
	\label{fig:decay}
\end{figure}

\begin{figure*}[t]
	\includegraphics[width=\textwidth]{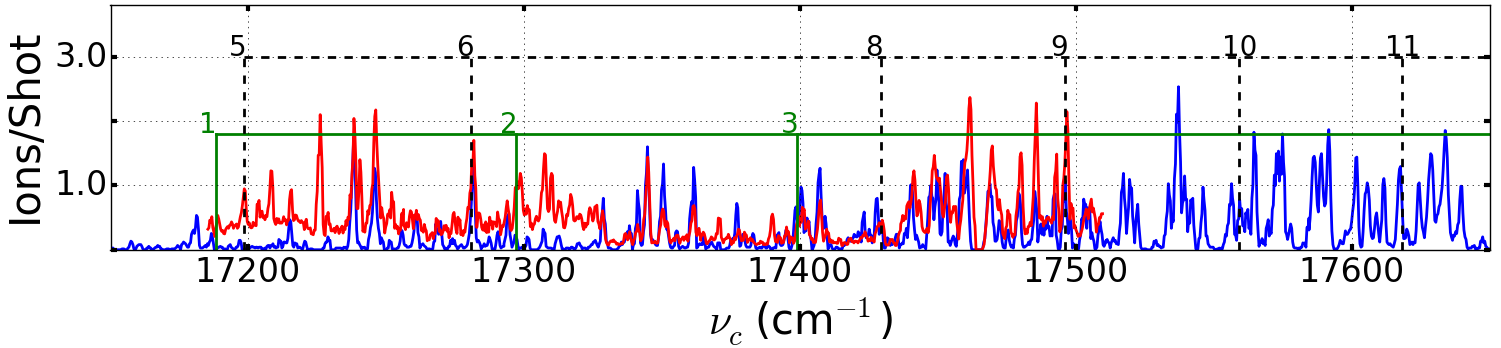}\\
	\caption{(Color online) REMPI (blue) and RE2PI (red) scan, the PA laser was locked to the $4(1) \ v=-16 \ J=1$ line.  Black numbers and dashed lines label $B \ ^1\Pi \ v^{\prime} \leftarrow v^{\prime \prime}=2$ transitions while green numbers and solid lines label $B \ ^1\Pi \ v^{\prime} \leftarrow v^{\prime \prime}=0$ transitions.  We have identified 92\% of the REMPI and RE2PI lines in this plot.  The RE2PI data is necessary because for deeply bound initial states ionization by REMPI is out of range energetically.  For example, $ v^{\prime}=1, \ 2, \ 3 \leftarrow v^{\prime \prime}=0$ and $v^{\prime}=5 \leftarrow v^{\prime \prime}=2$ on this plot would not be observable without RE2PI (even though there are nearby visible peaks in REMPI).}
	\label{fig:v0Progression}
\end{figure*} 

\section{Decay from 4(1) PA resonance}\label{sec:Decay}

In Sec.~\ref{sec:relativeDecay}, we showed that photoassociating Li and Rb atoms through the $|\Psi_- \rangle $ resonance leads with high probability to stable molecules, primarily in the $X ^1\Sigma^+$ electronic ground state.   Here we discuss the distribution of this $X ^1\Sigma^+$ electronic ground state population. 

We show in Fig.~\ref{fig:decay} the calculated Franck Condon factors for the spontaneous decay of the $B \: ^1 \Pi \ v^\prime=20$ ($4(1) \ v=-16$) to the various $v^{\prime \prime}$ levels of the $X ^1\Sigma^+$ state.  We calculated these FCFs using PECs from Ref.~\cite{ivanova2011x} and LEVEL 8.0~\cite{le2016level}.  This plot suggests that the $v^{\prime \prime} = 43$ level is the most highly populated, with a FCF of 0.35, and a secondary but still significant FCF of 0.13 for the $v^{\prime \prime} = 42$ level.  A broad pedestal of vibrational levels down to $v^{\prime \prime} = 2$ are also populated to a lesser degree.  Our observations here are in agreement with the calculated FCFs.  We used a second dye, LDS 698 which lases from $14050 - 15050$ cm$^{-1}$, to study population in $v^{\prime \prime}=38 - 45$.  We observed very strong population of $v^{\prime \prime} = 43$ and to a lesser extent $v^{\prime \prime}=42$.  Our RE2PI data on $v^{\prime \prime}=42$ and 43 will be presented elsewhere~\cite{stevensonCE16}.  Unfortunately, direct comparison of that data to the REMPI data presented in this work on deeply bound vibrational levels is difficult.  The two different dyes have different powers and modeling RE2PI is very difficult (modeling REMPI isn't any easier).  Additionally the intermediate state for that work, $C \ ^1\Sigma^+$, does not have a good theoretical model, so FCFs for the $C \ ^1\Sigma^+ \leftarrow X \ ^1\Sigma^+$ transition are unreliable, further complicating matters.  At this point in time, our best estimate for the production rate of $v^{\prime \prime} = 43$ is $R_{v^{\prime \prime}=43} = R^-_{PA} FCF_{43} (1.15 c^2) = 3 \times 10^5$ molecules/second.  The factor 1.15 represents the enhancement factor for bound state decay discussed in Sec.~\ref{sec:relativeDecay}.  This production rate is quite large and we expect that it could be increased further by increasing the PA power or MOT sizes.

To explore the deeply bound vibrational levels populated after PA through the 2(1) - 4(1) mixed states, we used REMPI and RE2PI with the R590 dye which lases from $17050 - 18150$ cm$^{-1}$.  We show a typical REMPI scan in Fig.~\ref{fig:v0Progression}.  In this spectrum, we scan $\nu_c$, the REMPI or RE2PI laser frequency, with the PA laser locked to the $4(1) \ v=-16 \ J=1$ line.  We have observed most of these REMPI lines previously in our studies of photoassociation through other vibrational lines of the $2(1)$ and $4(1)$ series~\cite{altaf2015formation,lorenz2014formation}.  Our current focus is finding population of low-lying vibrational levels.  For example, we have marked the series $B \: ^1 \Pi \ v^{\prime} \leftarrow v^{\prime \prime}=2$  transitions with black dashed vertical lines in Fig.~\ref{fig:v0Progression}.  The global maximum occurs at $\ v^{\prime }=14 \leftarrow v^{\prime \prime}=2$ (not shown).  The progression provides strong evidence of population of the $v^{\prime \prime}=2$ vibrational state.  

\begin{figure}
	\includegraphics[width=8.6cm]{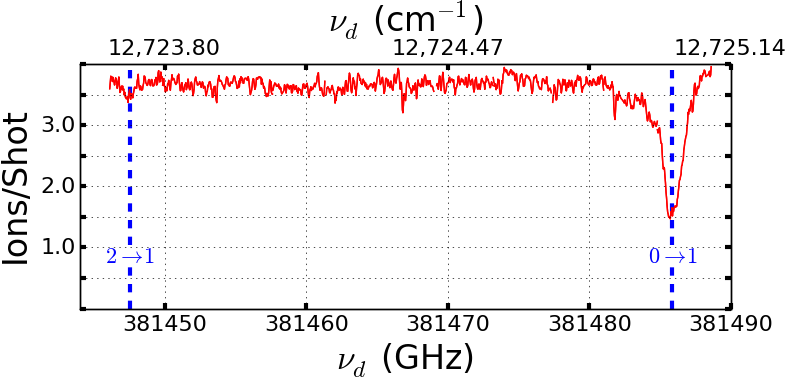}\\
	\caption{(Color online) Example of a depletion spectrum on the $ v^{\prime \prime}=2 $ state.  For this spectrum, the REMPI laser was tuned to the $B \ ^1\Pi \ v^{\prime}= 14 \leftarrow v^{\prime \prime}=2$ transition, the PA laser was locked to the $4(1) \ v=-16 \ J=1$ line, and $\nu_d$, the frequency of the depletion laser, was scanned.  Labeling is $J^{\prime \prime} \rightarrow J^\prime$.  Our observed $v^{\prime \prime}=2$ rotational constant, $B_{v^{\prime \prime}}$ = 6.38 GHz, matches the previous measurement~\cite{dutta2011laser}.}
	\label{fig:depletionExample}
\end{figure}

\begin{figure}
	\includegraphics[width=8.6cm]{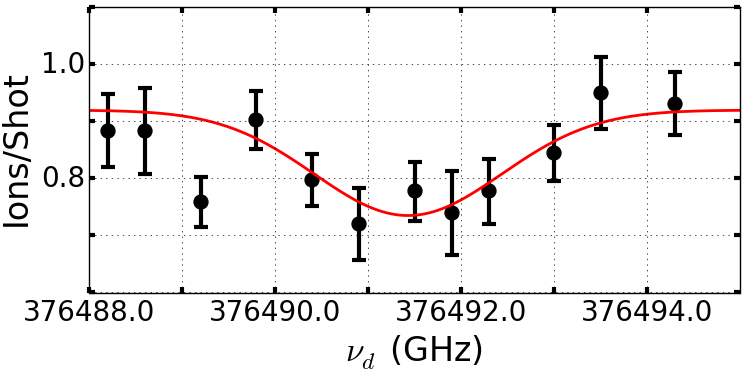}
	\caption{(Color online) Evidence of $X \ ^1\Sigma^+ \ v^{\prime \prime}=0$ population.  Depletion spectra of the $A \ ^1\Sigma^+\ v^{\prime}=10, J^{\prime}=1 \leftarrow v^{\prime \prime}=0$ transition.  The PA laser is locked to the $4(1) \ v=-16 \ J=1$ peak, the RE2PI laser was tuned to the $B \ ^1\Pi \ v^{\prime}=3 \leftarrow v^{\prime \prime}=0$ transition shown in Fig.~\ref{fig:v0Progression}. Here we extract a depletion of 0.19 (0.04) ions/shot.}
	\label{fig:v0num2}
\end{figure} 

In contrast, population of the $v^{\prime \prime}=0$ vibrational state is difficult to observe in the RE2PI spectra of Fig.~\ref{fig:v0Progression}.  The $4(1) \ v=-16 \ J=1$ state decays weakly to the ground vibrational state, predicted to be around 100 molecules/second, compared to other vibration levels such as $v^{\prime \prime}=2$, predicted to be around $2 \times 10^4$ molecules/second.  Furthermore, the candidate lines in Fig.~\ref{fig:v0Progression} that might originate from the $v^{\prime \prime}=0$ vibrational state, labeled by green solid lines, are obscured by nearby stronger lines.  For example, the $B \ ^1\Pi \ v^{\prime}=2 \leftarrow v^{\prime \prime}=0$ transition is much weaker than it looks on Fig.~\ref{fig:v0Progression} because it is right next to a much stronger $v^{\prime \prime}=4$ RE2PI transition.  In order to identify these weak lines we have employed a form of depletion spectroscopy~\cite{wang2007rotationally}, in which we introduce a cw laser beam tuned to a $A \ ^1\Sigma^+ \ v^{\prime} \leftarrow X \ ^1\Sigma^+ \ v^{\prime \prime}$ transition~\cite{stevensonCE16}.  For these measurements, we use the 150 mW ECDL to photoassociate the molecules, and the more tunable Ti:Sapphire laser to drive the depletion transition.  The PA and depletion beams copropagate, and are focused to a 200 $\mu$m diameter spot size in the MOT region.  To benchmark this technique, we first applied it to a $v^{\prime \prime}=2$ resonance with the result shown in Fig.~\ref{fig:depletionExample}.  For this spectrum, we tuned the REMPI laser frequency to the $B \ ^1\Pi \ v^{\prime}=14 \leftarrow v^{\prime \prime}=2$ transition, and tuned $\nu_d$, the depletion laser frequency, across the $A \ ^1\Sigma^+\ v^{\prime}=15 \leftarrow v^{\prime \prime}=2$ transition.  When on a depletion resonance, the initial state population available for ionization is transferred into a state not active in the REMPI process, thereby diminishing the REMPI signal.  Compared to REMPI, depletion spectra are sparse and have narrow peak widths, in this case $\sim$1 GHz, a typical linewidth for this type of measurement at this intensity~\cite{deiglmayr2008formation}.
  
In Fig.~\ref{fig:v0num2}, we show depletion of a $v^{\prime \prime}=0$ line.  For this depletion spectrum, we tuned the RE2PI laser to the $B \ ^1\Pi \ v^{\prime}=3 \leftarrow v^{\prime \prime}=0$ transition, and varied $\nu_d$, the depletion laser frequency, through the $A \ ^1\Sigma^+ \ v^{\prime}=10, \ J^{\prime}=1 \leftarrow v^{\prime \prime}=0, \ J^{\prime \prime}=0$ transition.  Each data point is the average of 10 measurements of the total ion count accumulated over 100 laser pulses, and error bars show the 1$\sigma$ standard deviation.  We extracted a depletion depth of 0.19 (0.04) ions/shot.  The data in Fig.~\ref{fig:v0num2} clearly shows population in the ground state $v^{\prime \prime}=0$ vibrational level.

We have two ways to estimate the $v^{\prime \prime} = 0 \ J^{\prime \prime} = 0$ production rate.  First, we measured the PA rate and calculated the branching ratio.  This gives us a $v^{\prime \prime} = 0 \ J^{\prime \prime} = 0$ production rate of 100 molecules/second.  Second, we measured the cold ion production rate in REMPI.  To get the cold molecule production rate: $R_{CM} = \frac{N}{\tau \epsilon_d P_{ion}}$, we estimate $\tau$, the transit time in the REMPI beam, to be 10 ms, $\epsilon_d$, the detector efficiency, to be about 50\% and $P_{ion}$, the ionization probability, to be about 5\% which gives us a generation rate of 600 molecules/second.  Within expected uncertainties, these two estimates of the production rate agree and we report the average, 300 molecules/second.  

In comparison to other experiments which generate $v^{\prime \prime} = 0 \ J^{\prime \prime} = 0$ by spontaneous decay, this rate is competitive.  In KRb, an impressive 5000 molecules/second (in an unspecified rotational distribution) can be produced~\cite{banerjee2012direct}, however the dipole moment, 0.57 Debye (D)~\cite{ni2008high} is very small compared to 4.1 D for LiRb.  Likewise, RbCs can be produced at around 2000 molecules/second~\cite{bruzewicz2014continuous}, but RbCs also has a small dipole moment at 1.2 D~\cite{molony2014creation}.  LiCs is produced at 100 molecules/second~\cite{deiglmayr2008formation}, but has a larger dipole moment of around 5.5 D~\cite{deiglmayr2010permanent}.  Finally, NaCs is produced at 10$^4$ molecules/second~\cite{zabawa2011formation}, the highest rate, and has a large dipole moment at 4.6 D~\cite{aymar2005calculation}, but these are all in $J^{\prime \prime} = 1$.  Overall, LiRb has a good mix of a strong dipole moment and good generation rate.

\section{Outlook}

Our production rate of $v^{\prime \prime} = 43 \ J^{\prime \prime} = 0$ is nearly 3 orders of magnitude larger than our $v^{\prime \prime}=0 \ J^{\prime \prime}=0$ production rate, at $2 \times 10^5$ molecules/second.  This is not the highest production of a single ro-vibrational level among bi-alkali's (or even in LiRb), but $v^{\prime \prime} = 43$ is sufficiently bound that a STIRAP transfer to the ro-vibronic ground state will not be too difficult.  Here, we recommend using the $C \ ^1\Sigma^+ \ v^{\prime} = 22$ as the intermediate state based on calculated FCFs~\cite{facts2}.  The `up' laser for this transfer is at 730 nm and the `down' laser is at 516 nm, both of which are covered by commercially available diode lasers.  Additionally, this STIRAP transfer will be several orders of magnitude stronger than transfers used by teams like the JILA KRb team~\cite{ni2008high}.  We predict (based on FCFs) that the transition dipole moment for each leg will be around 0.1 $e a_0$.  This is a huge advantage because it reduces the requirement for stability of the two STIRAP lasers, reducing the experimental and complexity cost considerably.  With this type of STIRAP transfer, we conservatively estimate producing $1 \times 10^5$ ro-vibronic ground state molecules/second.

\section{Conclusion}

We have experimentally explored $2(1) - 4(1)$ mixing that allows PA to a deeply bound $4(1)$ state.  From the spontaneous decay of our $4(1) \ v=-16 \ J=1$ PA resonance, we present the first observation of $X \: ^1 \Sigma ^+ \ v^{\prime \prime}=0$ in LiRb produced at  $\sim 300$ molecules/second.  We have presented plans for increasing the ro-vibronic ground state production rate by three orders of magnitude.  Currently, work is under way in our lab to study bound-to-bound transitions in LiRb with depletion spectroscopy using this PA resonance.  We will use the spectroscopic data to increase our ground state generation rate and devise a plan to trap ground state molecules.

We are happy to acknowledge useful conversations with O. Dulieu and Jes\'{u}s P\'{e}rez-R\'{\i}os, and university support of this work through the Purdue OVPR AMO incentive grant.  We are grateful to S. Dutta for providing feedback on our manuscript and we would like to acknowledge the work done by S. Dutta, A. Altaf, and J. Lorenz in building the LiRb machine.


\begin{thebibliography}{35}
\bibitem[{fac({\natexlab{a}})}]{facts1}
	\bibinfo{note}{The designation $v=-16 $ of the $4(1)$ state is only
		approximate.}
\bibitem[{fac({\natexlab{b}})}]{facts1.5}
	\bibinfo{note}{For convenience, we will continue to label these states $4(1) \
		v=-16$ and $2(1) \ v=-5$, according to their major constituent.}
\bibitem[{fac({\natexlab{c}})}]{facts2}
	\bibinfo{note}{As previously stated, FCFs for the $C \ ^1\Sigma^+ \leftarrow X
		\ ^1\Sigma^+$ are unreliable, so this state may change by a few vibrational
		numbers but we expect it to be close to the optimum state.}
\end{thebibliography}

\end{document}